\documentclass[useAMS,usenatbib]{mnras}
\usepackage{graphicx}
\usepackage{xcolor}
\usepackage{amssymb,amsmath,bm}
\usepackage{hyperref}
\usepackage[export]{adjustbox}
\hypersetup{colorlinks=true,citecolor=blue}
\usepackage{multirow}

\newcommand{\bq}{\textbf{q}}
\newcommand{\bx}{\textbf{x}}

\newcommand{\bPsi}{\boldsymbol{\Psi}}

\def \dtq{\int d^3 \bq \ }

\def\kMpc{\, h \, {\rm Mpc}^{-1}}

\title[Simulations and symmetries]{Simulations and symmetries}

\author[Modi, Chen \& White]
{
Chirag Modi$^{1}$, 
Shi-Fan Chen$^{1}$,
Martin White$^{1,2}$
 \\~\\
\footnotesize
$^1$Dept.~Physics, University of California, Berkeley, CA 94720, USA\\
$^2$Lawrence Berkeley National Laboratory, 1 Cyclotron Road, Berkeley, CA 93720, USA\\
}

\begin{document}
\maketitle

\begin{abstract}
We investigate the range of applicability of a model for the real-space power spectrum based on N-body dynamics and a (quadratic) Lagrangian bias expansion.  This combination uses the highly accurate particle displacements that can be efficiently achieved by modern N-body methods with a symmetries-based bias expansion which describes the clustering of any tracer on large scales.   We show that at low redshifts, and for moderately biased tracers, the substitution of N-body-determined dynamics improves over an equivalent model using perturbation theory by more than a factor of two in scale, while at high redshifts and for highly biased tracers the gains are more modest.  This hybrid approach lends itself well to emulation.  By removing the need to identify halos and subhalos, and by not requiring any galaxy-formation-related parameters to be included, the emulation task is significantly simplified at the cost of modeling a more limited range in scale.
\end{abstract}

\section{Introduction}
\label{sec:intro}

The study of the form and evolution of the large-scale structure in the Universe is one of the most promising probes of cosmology and fundamental physics \citep{Weinberg13,Amendola18}.  One of the major difficulties in interpreting data from large-scale structure surveys is that we measure a biased tracer of the non-linear density perturbations (and, for some surveys, in redshift space).  The combination of non-linear evolution and the non-linear dependence of galaxy bias makes robust inferences difficult.

The non-linearity of the dark matter field does not itself pose insurmountable difficulties.  On quasi-linear scales perturbation theory provides an accurate solution \citep[see][for recent examples]{VCW16,Ivanov19,DAmico19}.  Further, the evolution of dark matter particles under gravity from known initial conditions is a well posed numerical problem which can be solved with high accuracy and efficiency with modern N-body codes \citep{Springel05,Habib16,Garrison18}.  With care, percent level accuracy on the low order statistics of the density field can be obtained \citep{Heitmann08,Schneider16}, and interpolation formulae (`emulators') can be devised to provide predictions as a function of cosmological model \citep{Heitmann09,Heitmann10,Lawrence10,Zhai19,Knabenhans19,Wibking19}.

By contrast the behavior of the baryonic component, including hydrodynamics, star and black hole formation and feedback, remains a challenge.  Despite decades of progress in models, numerical algorithms, codes and computers a quantitative understanding of the translation from mass to light continues to elude us.  However, on sufficiently large scales all of these complexities can be parameterized by a series of numbers, the bias expansion, in a way that is informed by the symmetries of the underlying laws rather than the details of the specific processes that act \citep[see e.g.][for a recent review]{Desjacques18}.  
In detail, while the process that form and shape galaxies and other astrophysical objects are complex, all such objects arise from simple initial conditions acted upon by physical laws which obey well-known symmetries: for non-relativistic tracers these are the equivalence principle and translational, rotational and Galilean invariance.  This symmetries-based approach serves as a counterpoint to the ``halo model'' approach \citep[e.g.][]{Wechsler18}, which seeks to parameterize the manner in which galaxies inhabit halos of a given mass (and other properties).  While the latter offers us a fuller picture, which is more closely tied to the underlying physics, the former provides a fully flexible parameterization that captures the relevant effects on the large scales that dominate most cosmological inference (i.e.~on scales where the observed density field is still highly correlated with the early-Universe density field and the present day matter field).

A symmetries-based bias expansion is now quite common in theories which treat the dynamics perturbatively \citep{VCW16,Ivanov19,DAmico19, Colas19}, however the halo model approach is still more common in simulation-based approaches \citep[see e.g.][for recent examples]{Favole19,Wibking19,Zentner19,Zhai19}.  The purpose of this paper is to investigate the combination of the robust, symmetries-based bias expansion with the (well behaved) N-body solution to the dynamics.  Both the bias expansion and the N-body solution represent controlled approximations which can be made increasingly accurate given sufficient parameters and computational resources.  Further, the number of parameters and computational cost for a fixed accuracy can be lower than for many other schemes on the scales of relevance to next-generation large-scale structure surveys.

In this first paper we shall investigate how well a quadratic Lagrangian bias model, coupled with a ``full'' N-body dynamical model, can predict the real-space power spectrum of halos and mock galaxies.  Though the method can be straightforwardly extended to higher order in the bias (albiet with a large increase in the number of parameters that need to be included) and to configuration space, redshift space and higher order statistics, we focus first on the real-space power spectrum both because it is the simplest statistic and because it is of interest in interpreting projected statistics such as angular clustering and lensing (either of the CMB or of galaxies).  Recent related work on the accuracy of the Lagrangian bias expansion at the field level has appeared in \citet{Schmittfull19,Modi19b} and for Eulerian fields in \citet{Werner19}.

The outline of the paper is as follows: in the next section (\S\ref{sec:bias}) we introduce our (Lagrangian) bias expansion.  In \S\ref{sec:sims} we describe the N-body simulations which we use to compute our basis spectra and to test the performance of the model.  Our results are presented in \S\ref{sec:results}.  We present our conclusions and comment upon future directions in \S\ref{sec:conclusions}.

\section{The bias expansion}
\label{sec:bias}

In this paper we shall work within the context of Lagrangian bias, as formulated by \citet{Mat08}.  In such a prescription the (smoothed) initial distribution of tracers (e.g.~halos or galaxies) is obtained by ``weighting'' fluid elements by a functional, $F$, of the local initial conditions in the neighborhoods of their initial (or Lagrangian) positions, $\mathbf{q}$.  As long as we choose a sufficiently early time the fluctuations should be small and we can Taylor expand $F$.  We shall work to second order in the bias expansion and thus each particle in our N-body simulation will carry a weight
\begin{align}
     w(\bq) &= F\left[\delta_L(\mathbf{q}),\delta_L^2(\mathbf{q}),\nabla^2\delta_L(\mathbf{q}),s^2(\mathbf{q})\right] \nonumber \\
    &= 1 + b_1\delta_L(\mathbf{q}) + b_2\left( \delta_L^2(\mathbf{q})-\left\langle\delta_L^2(\mathbf{q})\right\rangle\right) \nonumber \\
    &+ b_s \left( s^2(\mathbf{q})-\left\langle s^2(\mathbf{q})\right\rangle\right) + b_{\nabla}\nabla^2\delta_L(\mathbf{q}). 
\label{eqn:weights}
\end{align}
where $s^2(\mathbf{q})$ is the (squared) shear field.
As is conventional, we define the linear overdensity, $\delta_L$, by its linearly-evolved value at the observed redshift, i.e.\ $\delta_L = D(z) \delta_{L,0}$, where $D(z)$ is the growth factor (normalized to unity at $z=0$).  Other conventions amount to a rescaling of the bias parameters, $b_i$.  The arguments of $F$ are all of the terms, to second order, allowed by symmetry and are to be interpreted as smoothed fields\footnote{Alternative bases for this expansion are possible, and sometimes used in the literature.  A change of basis would simply lead to a linear mixing of the bias parameters and would not fundamentally change our conclusions.}.

In general, the bias expansion quantifies the local response of the galaxy overdensity to long-wavelength density perturbations and will not hold to arbitrarily small scales. To lowest order, the effects of smoothing, as well as any ``non-local'' behaviors, are captured by the derivative bias $b_\nabla$.  We shall use the `natural' smoothing of our simulation grids ($0.75\,h^{-1}$Mpc), and comment upon this later \citep[see also][]{Aviles18}.  Going to higher order in the bias expansion requires the addition of many more terms, with cubic order already doubling the number of coefficients \citep{Lazeyras18,Abidi18}. Unlike the quadratic biases ($b_2$ and $b_s$), many of the cubic bias parameters have been detected in simulations with only a marginal significance even in more massive halo samples than the ones we investigate in this paper. 
We note that our scheme corresponds to selectively resumming only dynamical nonlinearities in the galaxy density field or, in the language of Eulerian bias, assuming values of Eulerian bias consistent with those generated by advection given nonzero $b_1$, $b_2$ and $b_s$.

The biased density field, $\delta^B(\mathbf{x})$, is then obtained by advecting the particles to their present day position, i.e.
\begin{equation}
    1 + \delta^B(\bx) = \dtq F(\bq)\ \delta_D(\bx - \bq - \bPsi(\bq)),
\label{eqn:advec}
\end{equation}  
where $\bPsi(\bq)$ denotes the displacement of fluid elements from their original (Lagrangian) positions to their final (Eulerian) positions.  We denote $\bPsi$ as a function of $\bq$ since this is common in the literature on Lagrangian perturbation theory and since in N-body simulations particles are often assigned ID numbers based on their initial positions.  We take the displacement, $\bPsi$, for each particle directly from the simulations.  Operationally $\delta^B$ can be easily computed by placing each particle onto a grid at its position at the time of interest (using the N-body position and possibly velocity) with a weight calculated from its initial position and the initial conditions according to Eq.~(\ref{eqn:weights}).

\begin{figure*}
\resizebox{\textwidth}{!}{\includegraphics{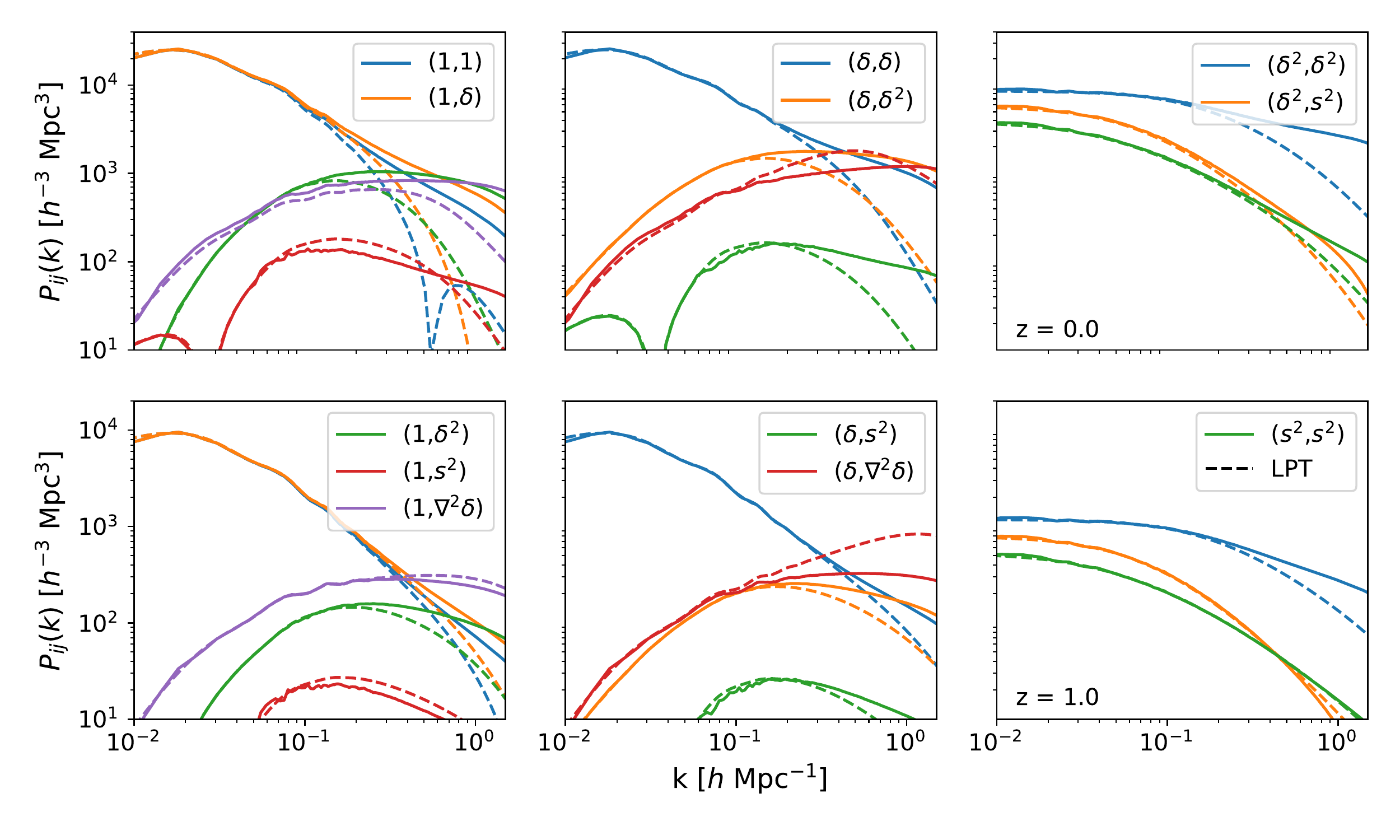}}
\caption{The 15 `basis' cross-spectra, $P_{ij}$, at $z=0$ (upper panels) and $z=1$ (lower panels). The halo and galaxy power spectra are formed from linear combinations of these spectra, as in Eq.~(\ref{eqn:power_spec_components}).  The matter and linear bias contributions ($P_{11}$, $P_{1,\delta}$ and $P_{\delta,\delta}$) dominate and are essentially degenerate on large scales, while differing at large $k$ where the other components also contribute.
The field $\nabla^2\delta$ has been multiplied by $10$ $h^{-2}$ Mpc$^{2}$ for ease of presentation.
}
\label{fig:components}
\end{figure*}

Within this formalism we can write halo power spectra as linear combinations of component cross spectra. Specifically, defining the component fields $\delta_i(\bx)$ as the initial fields $i = \{1, \delta_L, \delta_L^2, s^2_L, \nabla^2 \delta_L \}$ advected from $\mathbf{q}$ to $\mathbf{x}$ as in Eq.~(\ref{eqn:advec}), we have that the cross power spectrum between two biased tracers ($a$ and $b$) is given by 
\begin{equation}
    P^{ab}(k) = \sum_{i,j} F^a_i F^b_j P_{ij}(k) + P_{\rm SN}  ,
\label{eqn:power_spec_components}
\end{equation}
where $F^{a,b}$ are the coefficients in $w^a(\bq) = \sum_i F^a_i \delta_i(\bq)$ and, for example, $P_{\delta, \delta^2}$ is the cross spectrum between the advected linear density field and its square while $P_{11}$ is the (non-linear) matter power spectrum\footnote{We caution that our notation has e.g.~$P_{\delta,\delta^2}=P_{\delta^2,\delta}$ both contributing to $P^{ab}$.  The convention in perturbation theory calculations is often to absorb the factor of $2$ into the definition of $P_{\delta,\delta^2}$ and omit the second term.  We keep the symmetric form as it more naturally describes cross-spectra.}.
We also include a shot-noise term, $P_{\rm SN}$, to account for stochastic contributions to the halo field not accounted for by the bias expansion.
The extension of Eq.~(\ref{eqn:power_spec_components}) to multispectra is straightforward.
We emphasize that the 15 independent spectra, $P_{ij}$, can be individually computed from N-body simulations as described in the previous paragraph by weighting and advecting simulation particles, independently of the tracers in question.  Each of these spectra is a function only of the cosmology (and redshift), with all of the bias dependence for any tracer contained within the coefficients, $F$.  Avoiding the need to identify halos reduces the computational burden, both of finding the halos but also of sufficiently resolving them and possibly their histories, orientation, profiles and substructure.  The fact that $P(k)$ for all tracers (that can be described by quadratic bias) can be predicted from these $P_{ij}$ using Eq.~(\ref{eqn:power_spec_components}) means an emulator does not need to include any HOD-related parameters.

In the discussion above we have purposefully left out the effects of small-scale baryonic physics. This is because the bias expansion is only expected to be valid on scales where these baryonic effects -- for example due to AGN feedback or ionizing radiation -- are expected to be small \citep{Chisari19,Borrow19,vanDaalen19} and manifest as perturbative corrections $\propto k^2 P_L(k)$ to the power spectrum \citep{Lewandowski15,Schmidt17}.  Such corrections are nearly degenerate with contributions from derivative bias, $b_\nabla$ (e.g.\ the fitting function of \citet{vanDaalen19} is fit by $k^2P$ to one per cent on the scales where our bias model holds).  Indeed, the bias expansion itself would not be perturbative on scales where such baryonic effects are large.  On larger scales, baryons can also affect galaxy power spectra through primordial relative density and velocity perturbations \citep{Yoo11,Blazek16,Schmidt16,Chen19,Barreira19}. These effects are small and, while they are nondegenerate with contributions from our model, can be easily included at lowest order in perturbation theory. The inclusion of massive neutrinos is analogous, for light neutrinos.

\section{N-body simulations}
\label{sec:sims}

To investigate the performance of our quadratic bias model we make use of N-body simulations run for this purpose with the FastPM code \citep{Feng16}.  The FastPM code uses a relatively low resolution particle mesh algorithm with large, global timesteps to evolve particles and thus does not provide accurate predictions for the profiles or substructure in halos.  However, it does produce halo catalogs which are close to those produced by a more traditional N-body code \citep{Feng16,Ding18,Modi19a,Dai19}.  Since our purpose here is not to provide a percent level accurate prediction for a wide range of cosmologies but rather to test the performance of the bias model, any residual inaccuracy in the evolution should not be a concern: we aim to predict the clustering of halos and mock galaxies in the FastPM simulations using the particle dynamics generated by FastPM.

We ran 10 simulations, each of the same cosmology but differing in the random number seed used to generate the (Gaussian) initial conditions.  Each simulation employed $2048^3$ particles within a cubic, periodic box of side $1.536\,h^{-1}$Gpc, with forty time steps between redshifts z = 9 and 0 and snapshots output between $z = 3-0$.  The forces were computed on a $4096^3$ grid (i.e.~$B=2$). The simulations all assume a flat $\Lambda$CDM cosmology consistent with \citet{Planck18-I} ($\Omega_m = 0.309167, \Omega_b h^2 = 0.02247, \sigma_8 = 0.822, h = 0.677$).

We extract the particle data, and the friends-of-friends halo catalogs, from the outputs at $z=2$, $1$, $0.5$ and $0$.  We also use the initial conditions (at $z=9$), from which we generate the weights for each particle (Eq.~\ref{eqn:weights}).  Each particle is assigned a unique ID number to allow it to be tracked across outputs, and we compute the displacements simply by matching the initial and final positions for each particle.  We compute the weights from the initial conditions on a $2048^3$ grid corresponding to a $0.75\,h^{-1}$Mpc cell size.  We use cloud-in-cell interpolation of the particles onto the grid and of the weights onto the particles so this cell size forms a natural smoothing scale for our Lagrangian quantities.  That the cells are not $\ll 1\,h^{-1}$Mpc will affect the range over which we can expect to obtain good results, but we felt $0.75\,h^{-1}$Mpc was a good compromise between efficiency and convergence. We caution, however, that all Lagrangian weights are not created equal: while the linear weights are smoothed much like the matter and halo fields, quadratic weights like $\delta^2$ and $s^2$ are squares of smoothed fields which contain two factors of the window function, making them more susceptible to grid-size numerics.  We compute the component spectra using the NbodyKit software \citep{Hand18} using FFTs on $2048^3$ grids at the desired output time, with particles assigned to the grid using cloud-in-cell interpolation.  We do not subtract a (Poisson) shot-noise component from the spectra, as this is included in our model (Eq.~\ref{eqn:power_spec_components}; in all cases we find a best-fit $P_{\rm SN}$ that is within twenty per cent---and typically just a few per cent--- of the Poisson prediction).

\begin{table}
    \centering
    \begin{tabular}{ccccc}
                      & \multicolumn{2}{c}{$z=0$} & \multicolumn{2}{c}{$z=1$} \\
         $\log_{10}M$ & $\bar{n}$ & $b$ & $\bar{n}$ & $b$ \\  \hline
         (12.0,12.5)  &    24.3     & 0.80  &   23.7      & 1.30  \\
         (12.5,13.0)  &    9.5    & 0.89  &   7.9     & 1.69  \\
         (13.0,13.5)  &    3.6    & 1.10  &   2.2     & 2.36
    \end{tabular}
    \caption{Properties of the halo samples used in this work.  Halo masses are in $h^{-1}M_\odot$ and number densities in $10^{-4}\,h^{3}{\rm Mpc}^{-3}$.  The large-scale bias, $b$, is quoted as an Eulerian bias and is related to our Lagrangian bias, $b_1$, via $b=1+b_1$.}
    \label{tab:halo_samples}
\end{table}

We are interested in how well we can predict the real-space power spectra of (massive) halos and mock galaxies using our Lagrangian bias model.  Our focus will be $M>10^{12}\,h^{-1}M_\odot$ halos for two reasons.  First, these halos are better resolved allowing more accurate comparison with our theoretical model.  Second, these halos have higher and more scale-dependent bias, particularly at higher $z$, and so provide a stronger test of our model.  We consider three mass bins (see Table \ref{tab:halo_samples}) chosen to span a range of bias values while being well resolved and still having a high enough number density to permit good measurements of the power spectra: $12.0<\log_{10}M<12.5$, $12.5<\log_{10}M<13.0$ and $13.0<\log_{10}M<13.5$, with $M$ the halo mass measured in $h^{-1}M_\odot$.  We describe our model for mock galaxies, which occupy a range of halo masses and include both satellites and centrals, in \S\ref{ssec:galaxies}.

\section{Results}
\label{sec:results}

The Lagrangian prescription enables separate treatment of tracer bias and nonlinear dynamics. Section~\ref{sec:bias} describes a power spectrum model in which the latter are treated exactly (to simulation accuracy) while the former is treated perturbatively. By comparison, traditional approaches to perturbation theory (PT) treat both as effective expansions. As such, our approach can be expected to improve upon these calculations in the regime where the dynamics are no longer sufficiently captured by PT but the bias expansion remains valid, for example at low redshifts where dynamics become highly nonlinear but halos have relatively low biases. At high redshifts, where biases are large but dynamics essentially linear on most the scales of interest, our model should be valid over the same range of scales as traditional PT approaches.

The goal of this section is to investigate the range of scales over which our quadratic bias expansion is valid and useful. We proceed in two steps: in \S\ref{ssec:components}, we extract component spectra from the simulations and compare them to their predicitions in one-loop Lagrangian perturbation theory (LPT). Then, in \S\ref{ssec:halos}, we use the extracted component spectra to fit mass-limited halo power spectra and establish the scales over which the bias expansion is valid for various halo masses. Our model gains over traditional techniques in the regime where the dynamics are insufficiently captured by perturbation theory but the bias expansion remains valid.  We extend the comparison to mock galaxies, generated from a halo occupation distribution, in \S\ref{ssec:galaxies}.

\subsection{Component Spectra and Comparison to Perturbation Theory}
\label{ssec:components}

\begin{figure}
\resizebox{\columnwidth}{!}{\includegraphics{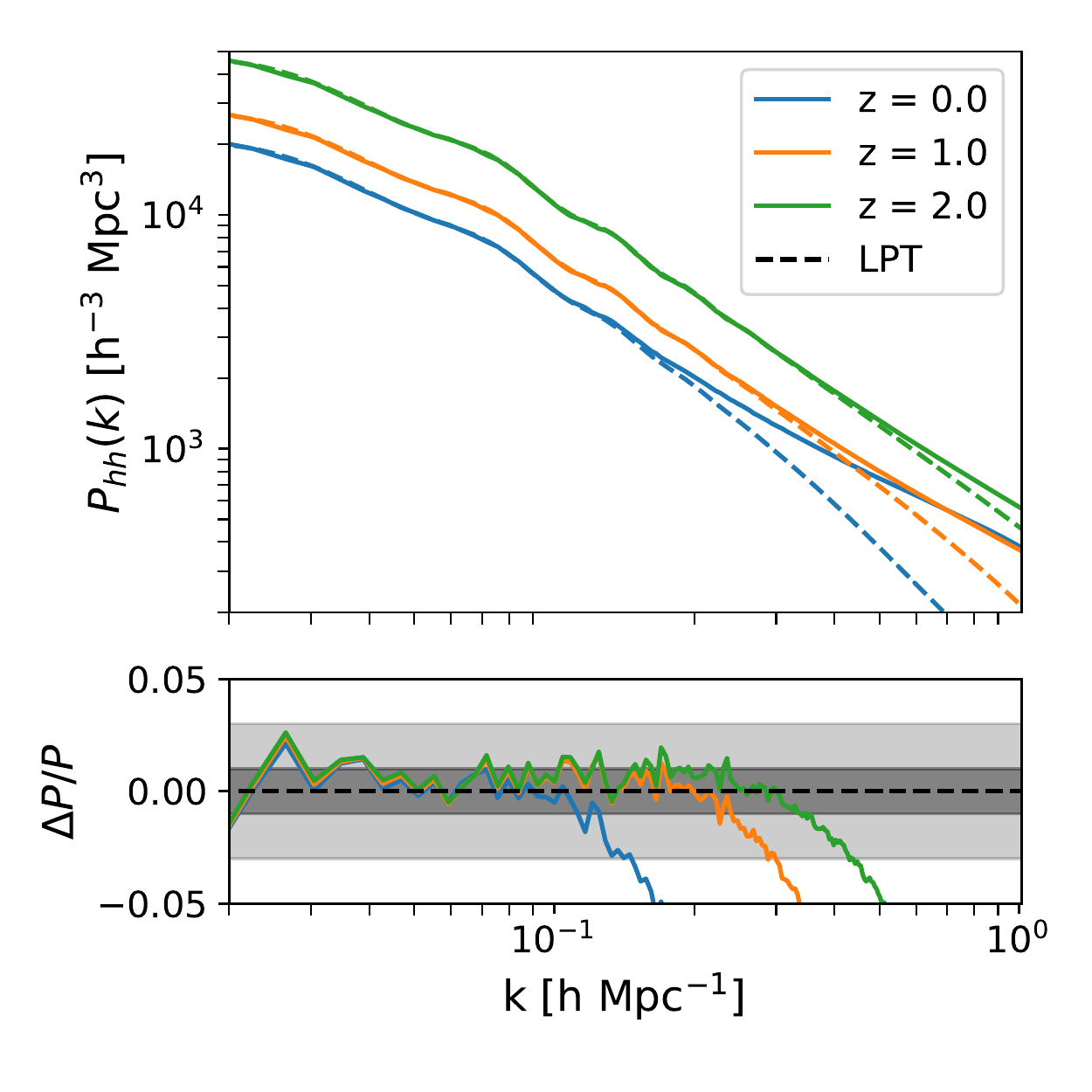}}
\caption{Comparison of halo autospectrum spectra predicted by our model and one-loop perturbation theory (LPT) for the same bias parameters. The latter matches our model on large scales but deviates towards large $k$ as perturbative dynamics breaks down, particularly at towards lower redshift.
}
\label{fig:compare_lpt}
\end{figure}

\begin{figure*}
    \centering
    \resizebox{\textwidth}{!}{\includegraphics{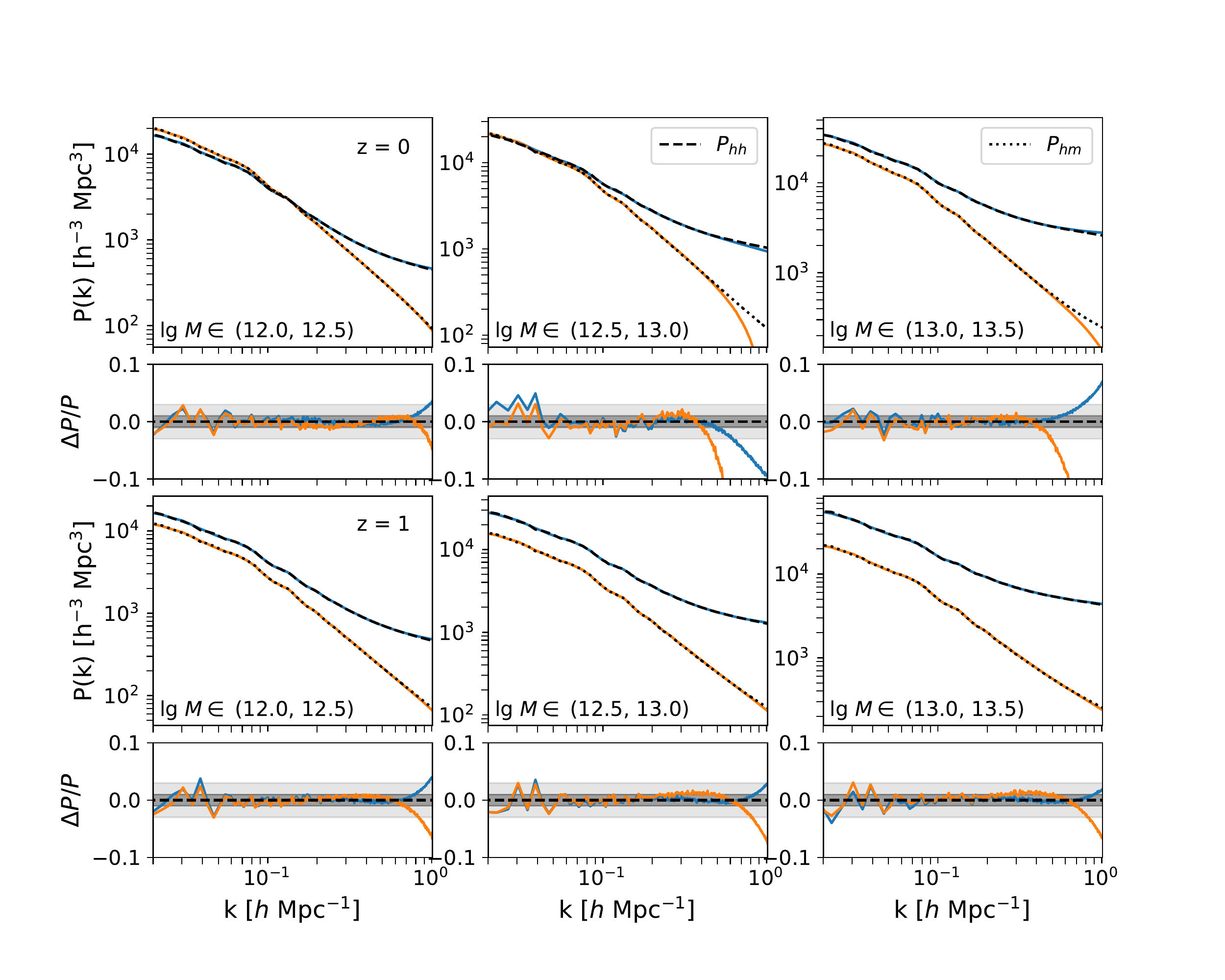}}
    \caption{Halo auto-spectra (dashed) and halo-matter cross-spectra (dotted) for our three halo samples (Left: $12.0<\log_{10}M<12.5$, Middle: $12.5<\log_{10}M<13.0$ and Right: $13.0<\log_{10}M<13.5$) at $z=0$ (top) and $z=1$ (bottom).  Black lines show the N-body spectra while the colored line shows the best-fit model of Eq.~(\protect\ref{eqn:power_spec_components}).  For each combination we show both the full spectra and the fractional error as a function of $k$. The gray lighter and darker shaded regions show 3 and 1 percent errors, respectively.
    }
\label{fig:halofits}
\end{figure*}

Figure~\ref{fig:components} shows the cross-spectra between the advected bias components, extracted from the simulations as described in \S\ref{sec:bias} and averaged over all ten simulation boxes, at redshifts $z = 0$ and 1\footnote{A similar plot appeared in Fig.~7 of \cite{Abidi18}, who compared cross-spectra of cubic fields to two-loop standard perturbation theory.}. We note that the cross spectra between linear and quadratic initial fields (e.g.\ $P_{\delta,\delta^2}$) are particularly noisy since their variance includes contributions cubic in the linear spectrum (e.g.\ $\sigma_{\delta, \delta^2}^2 \ni P_{\delta, \delta} P_{\delta^2, \delta^2} \sim \mathcal{O}(P_L^3)$) while their means are $\mathcal{O}(P_L^2)$ at lowest order, leading to a signal-to-noise ratio below unity.  We substitute the predictions of 1-loop LPT for these spectra at $k<0.08\,h\,{\rm Mpc}^{-1}$, where the theory is accurate but the N-body results very noisy\footnote{Since this component noise will also be present in any fitted data, given simulated volumes comparable to a given survey the summed model components will be no more noisy than the data even if some individual components have SNR less than unity.}.

Figure~\ref{fig:components} demonstrates that the matter and linear bias contributions ($P_{11}, P_{1,\delta},P_{\delta,\delta}$) dominate and are essentially degenerate on large scales, as expected. The dashed lines show the one-loop LPT predictions for these component spectra, which agree with the simulated component spectra on large scales but deviate on small scales where contributions due to quadratic and derivative bias also become significant, especially towards low redshifts\footnote{We have rescaled the $b_\nabla$ components to match $k^2 P_L(k)$ in physical units at large scales.}.

The dashed comparisons shown in Fig.~\ref{fig:components} were computed using ``traditional'' perturbation techniques; however, there has been much recent progress towards properly treating small-scale physics within the LPT framework using effective field theory techniques \citep{Por14,VlaWhiAvi15}, which must be included for a fair comparison with N-body simulations.  Figure~\ref{fig:compare_lpt} shows the predicted halo spectra within our model of quadratic bias plus N-body displacements (solid) compared to one-loop Lagrangian perturbation theory for values of bias $(b_1, b_2, b_s)$ that best fit the $12.5<\log_{10} M<13.0$ halos at $z=0, 1$ and 2. For simplicity we have not included nonzero derivative bias $b_\nabla$, but adjust a one-loop counterterm $\propto k^2P_L(k)$ for the LPT spectra to improve the agreement with simulation. In performing these fits we have adjusted the counterterm by eye to ensure good asymptotic behavior at large scales instead of maximizing the degree-of-fit over a wider range of $k$ in order to best show the domain of validity of LPT. At $z = 2$, one-loop perturbation theory shows good quantitative agreement with the the modeled N-body spectrum out to $k\simeq 0.5\kMpc$, while even with a relatively large counterterm it agrees with simulation only to $k\simeq 0.2\kMpc$ at $z = 0$.  These ranges-of-fit are consistent with the studies of the matter power spectrum within Lagrangian perturbation theory cited above and, roughly speaking, tell us when the nonlinear dynamics are no longer sufficiently described by perturbation theory (though some of the disagreement could also come from limited resolution in the simulations).  They suggest $P(k)$ cannot be fit beyond $k\Sigma\lesssim\mathcal{O}(1)$, where $\Sigma$ is the rms displacement of particles computed in linear theory, as would be expected on theoretical grounds.  We note that this comparison with LPT shares only one free parameter -- the counterterm -- with usual fits to N-body halo spectra, as the bias parameters are fixed.

Our conclusions are in good agreement with those of \citet{Munari17}, who showed that even if protohalo particles were properly identified in the initial conditions of a simulation using only perturbative displacements leads to poor prediction of $P(k)$ at non-linear scales.  Comparing\footnote{We thank E.~Castorina for emphasizing this point to us.} to their Fig.~3, it seems that the Lagrangian bias expansion does roughly as well as properly identifying protohalo particles in the initial conditions.

\subsection{Fitting halo spectra}
\label{ssec:halos}

Next we consider how well our model with N-body displacements predicts the (real space) halo auto-spectra and halo-matter cross-spectra for our three halo samples ($12.0<\log_{10}M<13.0$,  $12.5<\log_{10}M<13.0$ and $13.0<\log_{10}M<13.5$).  In each case we adjust both the 4 bias parameters plus the shot noise component to jointly fit the N-body halo autospectrum and halo-matter cross-spectrum.  We use a Gaussian approximation to the covariance of $P(k)$ to avoid noise in the error estimate from having only 10 independent realizations and consider the fits as a function of $k_{\rm max}$.  Once the $k$-modes become non-linear they also become increasingly correlated, and our error estimate thus gives too much weight to the high $k$ modes.  However, in this regime the noise is also very small and simply requiring our model to fit within 1 per cent is an effective strategy.

Figure \ref{fig:halofits} compares the halo auto-spectra and halo-matter cross-spectra for our two halo samples at $z=0$ and $z=1$ to the best-fit model of Eq.~(\ref{eqn:power_spec_components}).  The agreement for both statistics, with a common set of bias parameters, is excellent out to $k\simeq 0.6\,h\,{\rm Mpc}^{-1}$ for all three halo samples and both redshifts.  This substantially increases the range of fit at $z=0$, compared to the LPT described earlier, and corresponds to $kR_{\rm grid}\simeq 0.45$.  We have found that we could get even better agreement with only $P_{hh}(k)$, but at the cost of worsening the fit to $P_{hm}$.  This suggests that such good agreement with $P_{hh}$ is partially artificial, so we deal only with the joint fits in this paper.

There are several important features to note in Fig.~\ref{fig:halofits}.  First we see that the model is performing at the percent-level or better, and usually well within the errors of the simulation (visible as `noise' in the lines in the lower panels) at low and intermediate $k$, before a sudden shortfall of model power near $k\simeq 0.6\,h\,{\rm Mpc}^{-1}$ in the cross spectrum ($P_{hm}$). This rapid decline indicates that our component spectra are not well resolved at large $k$, which is to be expected given the finite size of the smoothing ($0.75\,h^{-1}$Mpc) we applied to estimate $\delta_L$, $\delta_L^2$ and $s^2$. This is especially true for the latter two which, as noted in \S\ref{sec:sims}, are particularly sensitive to smoothing. The auto spectrum ($P_{hh}$) is typically saturated by shot noise at $k\simeq 0.6\,h\,{\rm Mpc}^{-1}$ and therefore less sensitive to these effects.

Secondly, the model does better at $z=1$ than $z=0$, even though the values of the bias are higher.  This is because the linear growth factor drops by 40 per cent between $z = 0$ and $z = 1$, and for these samples the contributions from quadratic bias are relatively smaller at $z=1$ than $z=0$.  The improvement in the model performance is thus expected.

Finally, we note that in Fig.~\ref{fig:halofits} we haven't imposed any priors on the values of our bias parameters. While values of the derivative bias $b_\nabla$ will be sensitive to small-scale details such as smoothing and are therefore not expected to be universal, an extensive literature exists studying physical models for $b_1, b_2, b_s$ (see \S\ref{sec:intro} for references). To this end, we have checked that enforcing, to within a few per cent, the peak-background split relations between $(b_1,b_2)$ from \cite{ShethTormen99} (keeping $\nu$ as a free parameter) and values of $b_s$ from \cite{Abidi18} only degrades our fits at the few ($\sim 3$) per cent level in $P_{hh}$ and $P_{hm}$ and doesn't significantly alter the range of fit.

\begin{figure}
    \centering
    \includegraphics[width=0.45\textwidth]{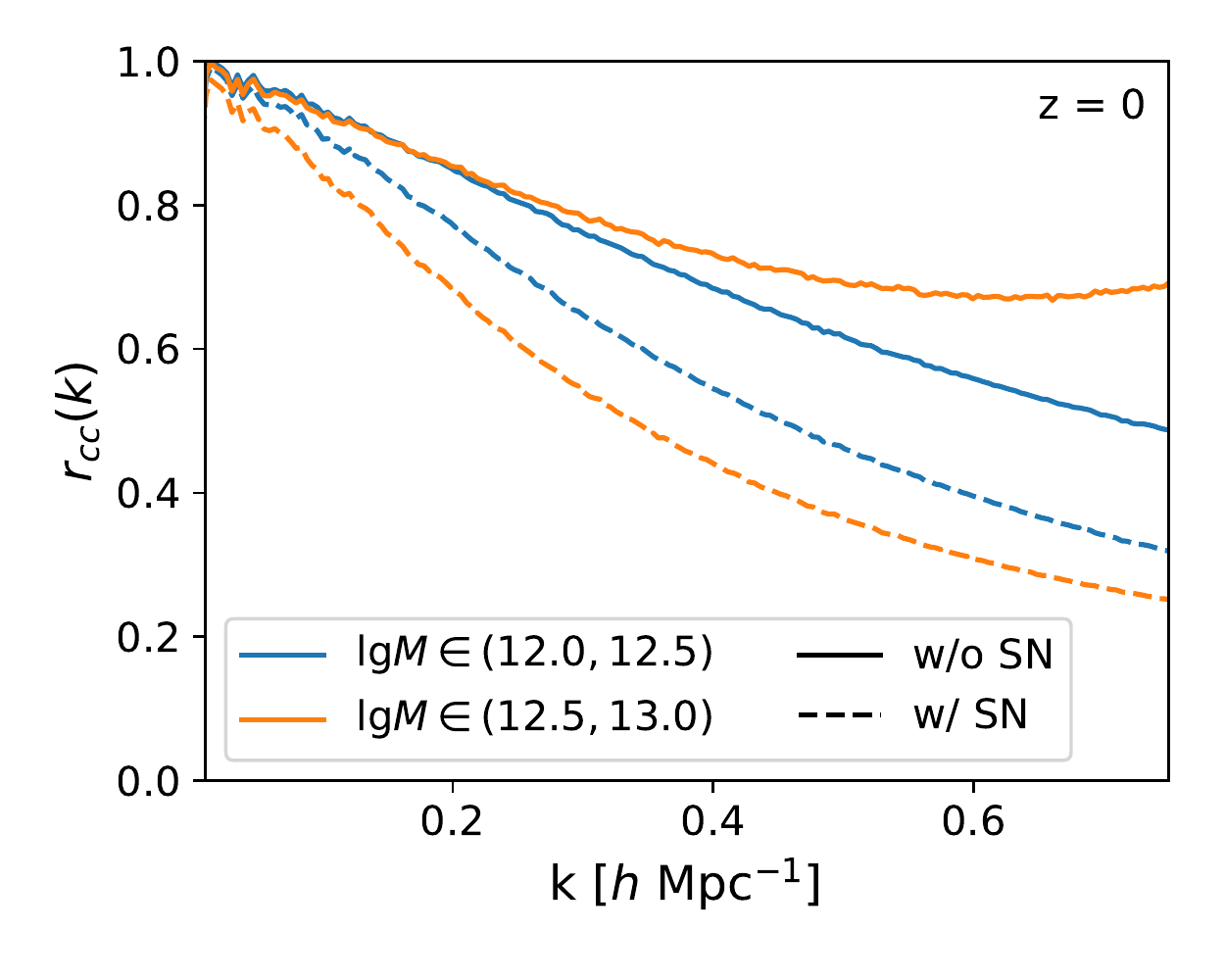}
    \caption{The scale-dependent matter-halo cross correlation coefficient, $r_{cc}(k)$, at $z = 0$ for mass bins $\log_{10}M \in (12.0,12.5)$ (blue) and $(12.5,13.0)$ (orange). The dashed lines show the ``true'' $r_{cc}$ while the solid lines show $r_{cc}$ computed without shot noise in the halo autospectrum, which gives a qualitative measure of the halo-matter decorrelation due to nonlinear dynamics and bias. In all cases the cross-correlation drops below one as the field goes non-linear and is less than 90 percent for most of the scales fit by our model.}
    \label{fig:rk}
\end{figure}

It is important to note that the quadratic bias model fits the auto- and cross-power spectra of the halo samples shown well into the quasi- or non-linear regime.  As modes become increasingly non-linear they also become increasingly correlated with each other and the halo field is much less correlated with the matter field or the initial density field. Figure \ref{fig:rk} shows the scale-dependent halo-matter cross-correlation coefficients
\begin{equation}
    r_{cc}(k) = \frac{P_{hm}(k)}{\sqrt{P_{hh}(k) P_{mm}(k)}}
\end{equation}
of two of our mass bins at $z = 0.$\footnote{We have avoided the highest mass bin with $\log_{10}M \in (13.0,13.5)$ as the halo power includes a significant contribution from shot noise at all scales.} We have computed $r_{cc}$ with and without the shot noise subtracted to better showcase the decorrelation due to nonlinear dynamics and bias, though we caution that strictly speaking the latter is the ``true'' cross-correlation coefficient. Nonetheless, in both cases $r_{cc}$ is at least ten per cent below unity across most of our fit range. For these reason the information content is substantially less than a simple mode-counting argument would suggest \citep[see e.g.][for discussion]{VN19,Wadekar19}.  It is also at these smaller scales that scale-dependent bias and complex physics involving the baryonic components becomes relevant, potentially requiring many more parameters to model faithfully.  Furthermore, most large-scale structure surveys are designed so that shot noise becomes comparable to the clustering signal near the non-linear scale, which further limits the information available from high $k$ modes.  Bearing all of this in mind, the performance of the quadratic bias model demonstrated above is likely to be sufficient for many science goals and we have not attempted to further improve it.

Figures \ref{fig:compare_lpt} and \ref{fig:halofits} demonstrate that, at low redshift, the perturbative dynamics breaks down before the quadratic bias model.  As we move to higher redshifts, and more biased tracers, the limitations imposed by perturbative dynamics become less severe and eventually we expect the bias model to become more limiting than the inaccuracies in the perturbative dynamics.  We have not investigated this limit.

\subsection{Fitting galaxy spectra}
\label{ssec:galaxies}

As a final test we fit to a mock galaxy sample, generated from our simulations by populating halos using a simple halo occupation distribution.  Specifically we assume the now-standard form \citep{Zheng05}
\begin{equation}
    \left\langle N_{\rm cen}\right\rangle(M_h) =
    \frac{1}{2}\left\{ 1 + \mathrm{erf}\left[\frac{\mathrm{lg}M/M_{\rm min}}{\sigma}\right] \right\}
\end{equation}
and
\begin{equation}
    \left\langle N_{\rm sat}\right\rangle(M_h) = \Theta(M_h - M_{\rm min})
    \left( \frac{M_h-M_{\rm min}}{M_1}\right)^{\alpha}
\end{equation}
For each halo in the simulation we draw a Poisson number of satellites and either 0 or 1 centrals.  The centrals are placed at the halo centers while the satellites are placed assuming an NFW profile \citep{NFW} dependent only on radius.

\begin{figure}
    \centering
    \resizebox{\columnwidth}{!}{\includegraphics{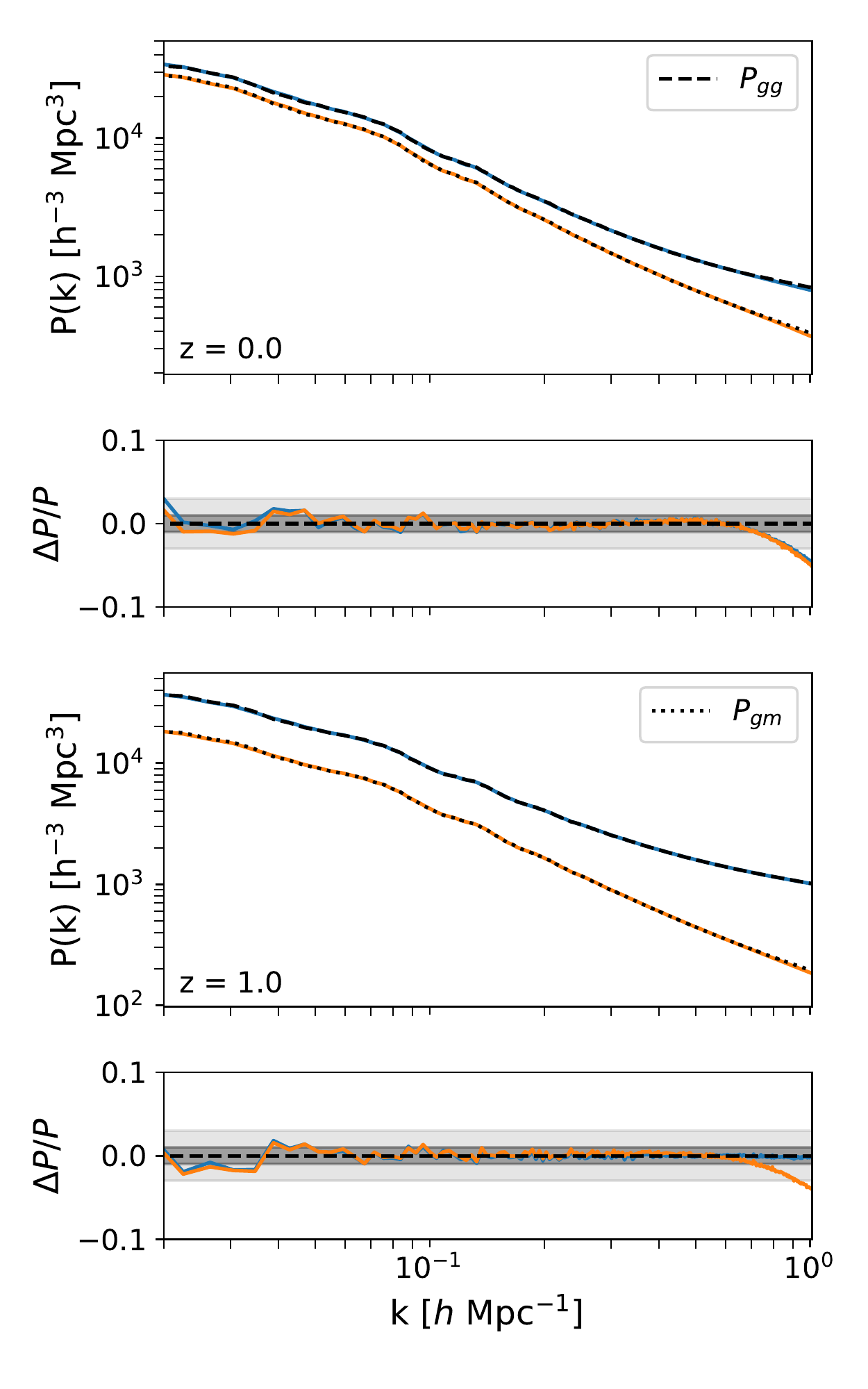}}
    \caption{Comparison of the auto- and cross-spectra for samples of mock galaxies, generated from the simulations using a halo occupation distribution at $z = 0$ (top) and $z = 1$(bottom). The blue and orange curves show the fits from our model for the galaxy autospectrum (dashed) and galaxy-matter cross spectrum (dotted), respectively. The model performance is qualitatively similar for our mock galaxies and halo samples.}
    \label{fig:galaxies}
\end{figure}

Figure \ref{fig:galaxies} shows $P_{gg}$ and $P_{gm}$ for a `galaxy' sample with $M_{\rm min}=10^{12.5}\,h^{-1}M_\odot$, $M_1=20\,M_{\rm min}$, $\sigma=0.2\,$dex and $\alpha=0.9$.  These are chosen to be similar to HODs found for magnitude limited samples of galaxies, though none of our conclusions depend upon the exact values of these parameters. For reference, our HOD parameters correspond to satellite fractions of $f_{\rm sat} = 0.18$ and  $0.1$ at $z = 0$ and $1$, respectively.

The results are very similar to those shown in Fig.~\ref{fig:halofits}.  The Lagrangian bias model fits the auto- and cross-spectra of our mock galaxies, simultaneously, within 3 per cent out to $k\simeq 0.6\,h\,{\rm Mpc}^{-1}$ for $0\le z\le 1$ (Fig.~\ref{fig:galaxies}).  This would be sufficient to model the angular clustering of galaxies in photometric surveys, galaxy-galaxy lensing or the cross-correlation of galaxies with CMB lensing out to angular multipole $\ell\approx k_{\rm max}\chi$ where $\chi$ is the characteristic distance to the objects in question.  Assuming $k_{\rm max}=0.6\,h\,{\rm Mpc}^{-1}$ and $\chi\approx 1.3\,h^{-1}$Gpc ($z=0.5$) gives $\ell_{\rm max}\simeq 800$ or $\ell_{\rm max}>10^3$ for $z>0.7$.  Beyond this $\ell_{\rm max}$ the errors grow, but smoothly rather than dramatically.  It is on these smaller scales that we expect contributions from baryonic physics to become increasingly important.

\subsection{Common bias model}

It is also instructive to compare our approach to the commonly assumed approximation of a constant or scale-dependent bias times the non-linear matter power spectrum.  Specifically we test the model
\begin{align}
    P_{hm} &= \left[ b_0' + b_1' k + b_2' k^2 \right]^{\hphantom{1}} P_m(k)
    \label{eqn:simple_model_hm} \\
    P_{hh} &= \left[ b_0' + b_1' k + b_2' k^2 \right]^2 P_m(k)  + P_{SN} \label{eqn:simple_model_hh}
\end{align}
with three bias and one constant shot noise parameter. The parameter $b_0'$ denotes a scale-independent bias, and is the most widely used model for galaxy or halo bias.  The $b_2'$ term describes a correction due to peaks theory \citep{Desjacques18} and has been used in modeling data \citep[e.g.][]{Giusarma18}.  The term $b_1'\, k$ has no theoretical justification and is included merely because we noted that it improved the fit.  We use the N-body determined $P_m(k)$ in Eqs.~(\ref{eqn:simple_model_hm}, \ref{eqn:simple_model_hh}) as we found the HaloFit model \citep{Hamilton91,PD96,Smith03,Takahashi12,Mead15,Mead16} was not as accurate and we wished to provide the most fair comparison. 

\begin{figure}
    \centering
    \includegraphics[width=0.45\textwidth]{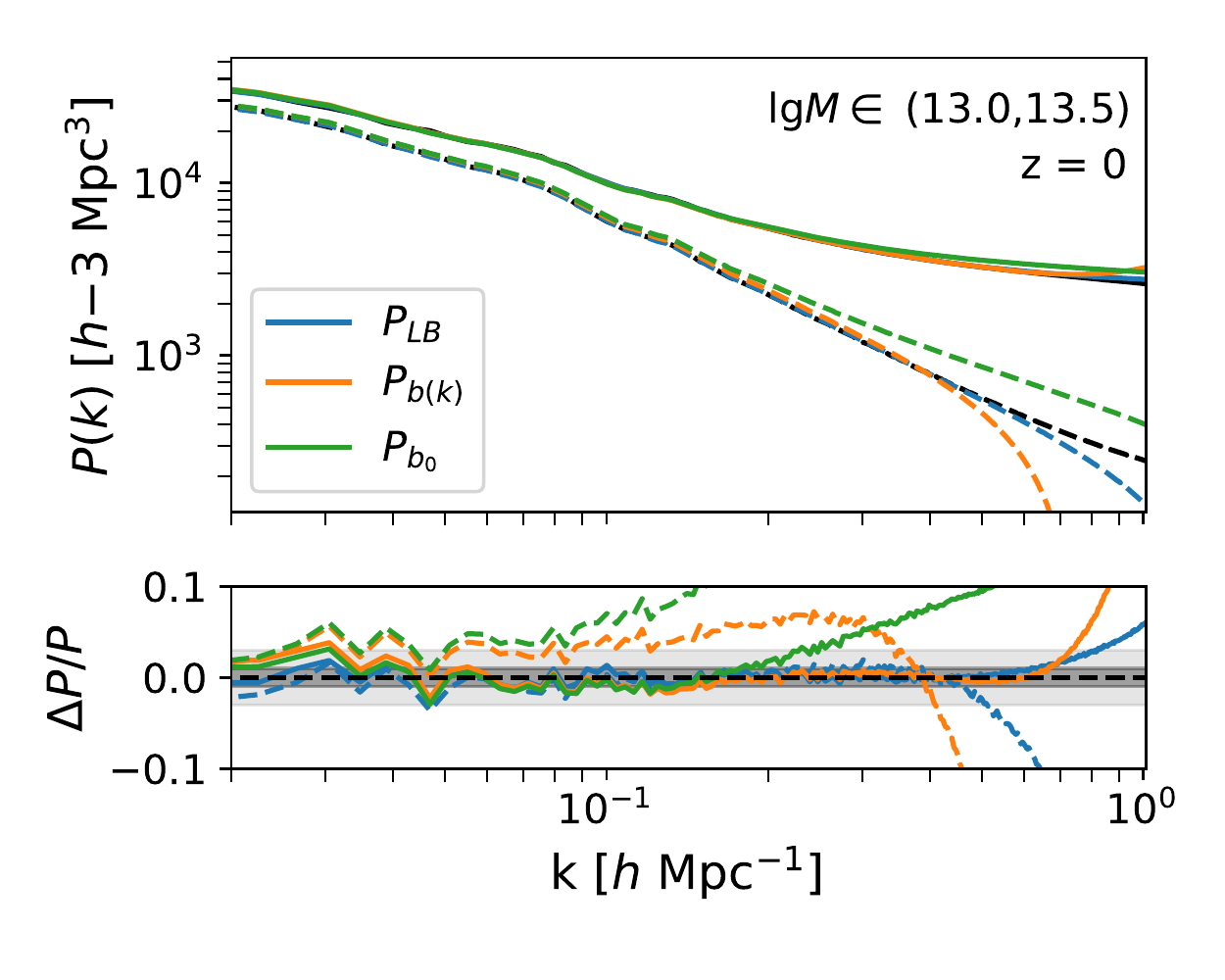}
    \caption{A comparison of our Lagrangian bias model with the model of Eqs.~(\ref{eqn:simple_model_hm}, \ref{eqn:simple_model_hh}) and the benchmark linear bias model.  Solid lines show the fits of each model to the halo-halo autospectrum, while dashed lines show fits to the halo-matter cross spectrum. The linear bias model only fits the data on the largest scales. While the scale-dependent bias model can be made to fit the autospectrum, only our model fits both auto- and cross-spectra with a consistent set of parameters.
    }
\label{fig:simple_model}
\end{figure}

Note the assumption above that the prefactor of the halo-halo auto-correlation is the square of the prefactor in the halo-mass cross-spectrum.  This is equivalent to the assumption that the halo and matter field have cross-correlation coefficient $r_{cc}\approx 1$.  However, this assumption increasingly breaks down as dynamics and bias become nonlinear at low redshift and high mass (Fig.~\ref{fig:rk}; see also \citealt{Modi17b,Wilson19}). The model of Eq.~(\ref{eqn:power_spec_components}) allows us to relax the assumption that $r_{cc} = 1$.

Figure \ref{fig:simple_model} shows the results at $z=0$ for the 4-parameter model (Eqs.~\ref{eqn:simple_model_hm}, \ref{eqn:simple_model_hh}) on the halo sample with $13.0<\log_{10}M<13.5$. We have chosen this redshift and mass bin as it illustrates dynamics and biasing at their most nonlinear, though other choices yield qualitatively similar results. As a reference, we also consider the case of constant bias (only $b_0\ne 0$ above).
While the Lagrangian bias model provides a good fit to both spectra simultaneously, as we have seen previously, this is not true of Eqs.~(\ref{eqn:simple_model_hm}, \ref{eqn:simple_model_hh}).  We have chosen to adjust the parameters in $b(k)$ to predict $P_{hh}$ on quasi-linear scales as in observations $P_{hh}$ would most likely have the highest signal to noise ratio.  The freedom inherent in the quadratic function, $b_0'+b_1' k+b_2' k^2$, allows us to fit $P_{hh}$ well up to $k\approx 0.8\,h\,{\rm Mpc}^{-1}$, comparable to our Lagrangian bias model.  However the form preferred by $P_{hh}$ provides a very bad fit to $P_{hm}$ at intermediate to high $k$, as can most easily be seen in the lower panel of Fig.~\ref{fig:simple_model}.  This leads to a significant misestimate of $P_{hm}$, which would translate into errors in the inferred large-scale bias and underlying matter clustering amplitude ($\sigma_8$).

Despite its ubiquity in analyses, the constant bias model does even more poorly.  The significant scale-dependent bias inherent in the clustering of this mock galaxy sample makes it impossible to fit both the auto- and cross-spectra except at the very largest scales, $k<0.1\,h\,{\rm Mpc}^{-1}$.  Inferences about cosmological parameters from using this model would be highly biased unless drastic scale cuts were employed.

\section{Conclusions}
\label{sec:conclusions}

\begin{figure}
    \centering
    \includegraphics[width=0.45\textwidth]{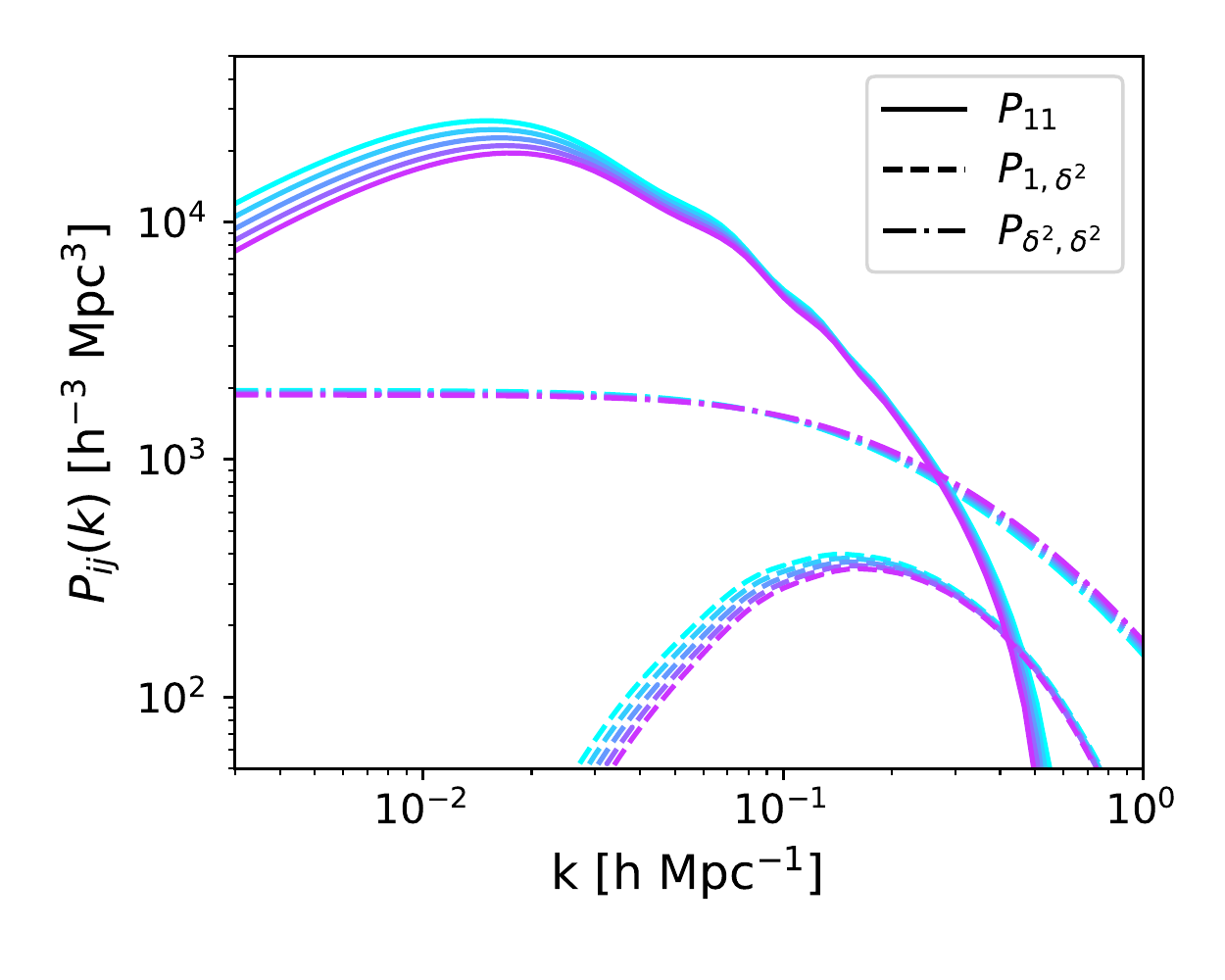}
    \caption{The cosmology dependence of the component spectra. Here we show three representative components: $P_{11}, P_{1,\delta^2}, P_{\delta^2,\delta^2}$ at $z = 0$ for values of $\Omega_m$ within ten percent of our fidicucial cosmology, with all other parameters kept fixed. For simplicity we have used 1-loop LPT as a proxy for the N-body spectra. The components vary smoothly with cosmology, with $P_{\delta^2, \delta^2}$ showing very little variation.  Critically, the component spectra change with cosmology at about the same rate as (or less than) the matter power spectrum, $P_{1,1}$.
    }
\label{fig:cosmo_dep}
\end{figure}

We have tested the performance of a power spectrum model for biased tracers based on a quadratic, Lagrangian bias expansion.  The model uses N-body simulations to compute the gravitational evolution of dark matter particles, but substitutes a 4-parameter bias model for the halo-based galaxy modeling more traditionally employed in simulations.  Both the dynamical model and bias expansion are theoretically well motivated, and the method places only modest requirements on the input simulations since it does not explicitly use properties of halos or subhalos.  This is an advantage given that properly resolving halos and subhalos is quite computationally demanding \citep{vandenBosch18,deRose19,Dai19} and complex halo occupations -- potentially including halo assembly information -- can be required in order to properly model samples selected by emission lines, color cuts or other complex selections \citep{Reid14,Favole17,Zhai17,Campbell18,Wechsler18,Favole19,Mansfield19,Wibking19,Zentner19,Zhai19}.  The approach combines methods from the `analytic' and `numerical' communities in a manner which plays to their relative strengths.

The Lagrangian bias model is quite accurate on large and intermediate scales.  We have showed that going to quadratic order in the bias expansion enables us to fit the (real space) auto- and cross-power spectra of halos and mock galaxies to a few per cent out to $k\simeq 0.6\,h\,{\rm Mpc}^{-1}$ for $0\le z\le 1$ (Figs.~\ref{fig:halofits}, \ref{fig:galaxies}).  To fit beyond this scale would require increasing the number of parameters (and component spectra) and calculating the $P_{ij}$ with higher resolution simulations.  However, this performance is already highly encouraging, as these scales provide the bulk of the information in many cosmological analyses.  Smaller scales tend to be non-linear and significantly affected by scale-dependent bias and baryonic effects.  The mode-coupling associated with non-linearity implies that there is less information about primordial physics in these modes than a simple mode-counting exercise would imply \citep[e.g.][]{VN19,Wadekar19} and the combination of non-linearity and baryonic effects means that such modes do not faithfully trace the primordial perturbations.  The many parameters needed to describe complex, scale-dependent effects can lead to degeneracies with cosmological parameters.  Furthermore, most large-scale structure surveys are designed so that shot noise becomes comparable to the clustering signal near the non-linear scale, which further limits the information available from high $k$ modes.  For these reasons, the performance of the quadratic bias model is likely to be sufficient for many science goals.

In this paper we have worked at fixed cosmology in order to focus on the range of applicability of the quadratic bias expansion.  While we intend to return to the problem of emulating the power spectrum for different cosmologies in future work, we comment here on the basic strategy.  The component spectra in Eq.~(\ref{eqn:power_spec_components}) vary with cosmology smoothly, with variations similar to the linear power spectrum. As an example, in Figure~\ref{fig:cosmo_dep} we have plotted variations in the component spectra when $\Omega_m$ is varied within $\pm 10$ per cent from our fiducial cosmology; leading order terms like the matter power spectrum $P_{11}$ vary like the linear power spectrum, while the component spectra vary smoothly by similar factors or, in the case of $P_{\delta^2,\delta^2}$, significantly less.  The variations with other cosmological parameters are qualitatively similar. Thus the same techniques that have been used to emulate matter power spectra will apply almost unchanged for emulating $P_{ij}$.  As shown in Fig.~\ref{fig:compare_lpt} we can use perturbative methods for the low $k$ part of the component spectra, which tends to be relatively noisy when estimated from simulations of computationally tractable volumes, and switch to N-body determined spectra at higher $k$.  Given a grid of N-body simulations spanning the cosmologies of interest standard Gaussian process regression, which has been successfully used for matter power spectrum interpolation \citep{Heitmann09,Heitmann10,Lawrence17,Knabenhans19, vanDaalen19}, can easily be used to predict each of the component spectra as a function of cosmology.  In a similar vein, the ratio of the N-body to perturbation theory spectra can be emulated rather than the spectra themselves, removing some of the cosmology dependence.  Since the perturbation theory spectra can be efficiently and accurately computed for any cosmology, this shouldn't significantly change the efficiency of the emulator.

An alternate emulation which also does not explicitly use properties of halos and subhalos was adopted by \citet{hzpt, Hand17b}. Those authors used Pade approximants to fit correction factors to perturbation theory or halo model inspired terms and then fit the coefficients as power laws in the relevant cosmological parameters. Such an approach could also be attempted with our component spectra, which are in large part relatively featureless and vary smoothly with parameters.

While we have chosen a Lagrangian bias expansion, a similar procedure could be followed using a complete set of Eulerian bias operators. However, we note that \citet{Schmittfull19, Modi19b} find that the Lagrangian scheme outperforms the Eulerian bias expansion for a wide range of halo masses, redshifts and weightings. Thus we do not expect it to improve over the prescription we have developed here.

In this paper our focus has been on the real-space power spectrum, of direct relevance to modeling photometric and lensing surveys, though one can extend the method to higher order functions, covariances and to redshift space.  For the latter, one can either model the contributions to $P(k,\mu)$ directly in simulations, or one can choose to model the real-space power spectrum and velocity moments and construct the redshift-space power spectrum from those components (see e.g.~\citealt{Hand17b} for a recent example and \citealt{VlaWhi19}  for a recent discussion of such methods for modeling redshift-space distortions).  We intend to return to this topic, and to the construction of an emulator, in future publications.

\vspace{0.2in}
The authors thank E.~Castorina, J.~Cohn, M.~Schmittfull and U.~Seljak for helpful comments on an earlier draft. S.C.~also thanks M.~Simonovic and Z.~Vlah for helpful discussions while this paper was being revised.
S.C.~is supported by the National Science Foundation Graduate Research Fellowship (Grant No.
DGE 1106400) and by the UC Berkeley Theoretical Astrophysics Center Astronomy and Astrophysics
Graduate Fellowship. M.W.~is supported by the U.S.~Department of Energy and by NSF grant number 1713791.
This research used resources of the National Energy Research Scientific Computing Center (NERSC), a U.S. Department of Energy Office of Science User Facility operated under Contract No. DE-AC02-05CH11231.
This work made extensive use of the NASA Astrophysics Data System and of the {\tt astro-ph} preprint archive at {\tt arXiv.org}. 

\bibliography{main}

\end{document}